\RequirePackage[colorlinks,allcolors=blue,pdfborder={0 0 0}]{hyperref}
\documentclass[draft]{agujournal2019}
\usepackage{url} 
\usepackage{soul}
\usepackage{amsmath}
\usepackage[separate-uncertainty=true, multi-part-units=single]{siunitx}
\usepackage[dvipsnames]{xcolor}
\sethlcolor{Salmon!15}

\AtBeginDocument{\renewcommand{\APACrefYearMonthDay}[3]{\BBOP{#1}\BBCP}}
\AtBeginDocument{}

\draftfalse

\journalname{}

\begin{document}

%
%

\title{Weak 21\textsuperscript{st}-century AMOC response to Greenland meltwater in a strongly eddying ocean model}

%
%



\authors{Oliver Mehling and Henk A. Dijkstra}

\affiliation{}{Institute for Marine and Atmospheric research Utrecht, Department of Physics, Utrecht University, Utrecht, The Netherlands}

\correspondingauthor{Oliver Mehling}{o.m.mehling@uu.nl}


\begin{keypoints}
\item The impact of increasing Greenland meltwater on the AMOC under climate change is studied with a strongly eddying ocean model
\item Meltwater induces an additional AMOC weakening of 0.6 $\pm$ 0.2 Sv by 2100, similar to values obtained at low resolution and in coupled models
\item Increasing stratification under global warming shapes the state-dependent AMOC response to Greenland meltwater
\end{keypoints}

%
%

%
%


\begin{abstract}
Climate models project that the Atlantic Meridional Overturning Circulation (AMOC) will weaken in the 21\textsuperscript{st} century, but the magnitude is highly uncertain. Some of this uncertainty is structural, as most climate models neglect increasing meltwater from the Greenland ice sheet and do not explicitly capture mesoscale ocean eddies. Here, we quantify the impact of Greenland meltwater on the AMOC until 2100 under SSP5-8.5 forcing for the first time in a strongly eddying (1/10° horizontal resolution) ocean model. The meltwater-induced additional AMOC weakening is small (\SI{0.6(2)}{Sv}) compared to the weakening due to warming alone, and similar at high and low resolution. The same meltwater would cause a stronger AMOC weakening under present-day climate conditions. We link both resolution-independence and state-dependence to large-scale controls of the AMOC. Our results demonstrate that the background ocean state is more important than resolution in determining how Greenland meltwater affects the AMOC.
\end{abstract}

\section*{Plain Language Summary}
Human-caused global warming is expected to cause a weakening of the Atlantic Meridional Overturning Circulation (AMOC) and increased melt of the Greenland ice sheet. However, most current climate models do not represent the interaction between these two systems. The question of how strongly Greenland melt will impact the AMOC in the 21\textsuperscript{st} century has already been studied in climate models with an ocean resolution of around 100 km. However, this is too coarse to explicitly capture mesoscale eddies, which could influence how meltwater propagates in the North Atlantic and how it affects the AMOC. Here, we quantify the impacts of Greenland meltwater on the AMOC until 2100 in a high-resolution (around 10 km horizontal resolution) eddy-rich ocean model. We find that the AMOC response to Greenland meltwater under global warming is similar at high and low resolution, and -- consistently with previous studies -- that it is small compared to the AMOC weakening driven by global warming alone. The same amount of meltwater also weakens the AMOC less strongly under global warming than under present-day climate conditions. This shows that the background ocean state is more important than resolution in determining how Greenland melt affects the AMOC.

%
%

\section{Introduction}
The Atlantic Meridional Overturning Circulation (AMOC) is expected to play an important role in shaping the climate response to future anthropogenic greenhouse gas emissions \cite{Liu2020,Bellomo2021,Bellomo2024}. While there is high confidence in climate model projections that the AMOC will weaken in the 21\textsuperscript{st} century \cite{Weijer2020}, the magnitude of this weakening is highly uncertain \cite{Fox-Kemper2021}. This is not only due to large inter-model spread in the simulated AMOC decline \cite{Reintges2017,Weijer2020,Bonan2025,Portmann2026}, but also due to the potential of nonlinear AMOC changes such as an abrupt AMOC collapse or tipping, the probability of which remains poorly quantified in climate models \cite{Dijkstra2026,Loriani2025}.

Based on evidence from current-generation climate models, the IPCC Sixth Assessment Report assessed with ``medium confidence'' that the AMOC ``will not experience an abrupt collapse before 2100'' \cite{Fox-Kemper2021}. The main reasons for assigning only ``medium confidence'' to this finding were structural deficits such as model biases related to AMOC stability \cite{Liu2017,Mecking2017,VanWesten2024b} and neglected meltwater influx from the Greenland Ice Sheet (GrIS).

The impact of realistic Greenland meltwater input on 21\textsuperscript{st}-century AMOC changes has been extensively studied in IPCC-class climate models with a typical ocean resolution of around 1° \cite<e.g.,>{Lenaerts2015,Bakker2016,Ackermann2020,Mehling2026}. These models show a moderate response of less than or around \SI{1}{Sv} ($\SI{1}{Sv} = \SI{e6}{\cubic\meter\per\second}$) meltwater-induced additional AMOC weakening until 2100 but do not explicitly capture mesoscale ocean eddies. It has been shown that, under historical climate conditions, resolving mesoscale eddies can have a potentially important effect on how Greenland meltwater affects the large-scale ocean circulation \cite{Weijer2012,Boning2016,Swingedouw2022,Martin2023,Martin-Martinez2025}. While 21\textsuperscript{st}-century climate projections have recently been carried out at eddy-rich (1/10° or higher) resolution \cite{Chang2020,Juling2021}, the effect of Greenland meltwater on the AMOC under 21\textsuperscript{st}-century climate change has so far not been separately assessed with an explicit representation of mesoscale eddies. To our knowledge, the only attempt in this direction has been by \citeA{Li2023a}; however, their simulations ended in 2050, focused on the Antarctic Bottom Water response to meltwater, and did not capture any CO$_2$-induced AMOC weakening.

In this study, we use a global ocean model at 1/10° resolution to quantify the 21\textsuperscript{st}-century AMOC response to Greenland meltwater under a strong global warming scenario. A high-end but physically plausible estimate of Greenland ice sheet runoff is derived from a fully coupled climate--ice sheet model simulation \cite{Muntjewerf2020}. Comparison with the low-resolution version of the same model and a set of previous, more idealized Greenland meltwater simulations under present-day conditions \cite{Weijer2012} allow disentangling the impacts of resolution and background climate state on the AMOC response to 21\textsuperscript{st}-century Greenland melt.

\section{Materials and Methods}

\subsection{Global ocean model}
Simulations are performed with version $2\alpha$ of the Parallel Ocean Program \cite<POP;>{Dukowicz1994}. The high-resolution global configuration of POP \cite{Maltrud2010,Weijer2012} (HR-POP hereafter) has a resolution of 1/10° using a tripolar grid in the horizontal and 42 levels in the vertical, with a top-layer thickness of \SI{10}{m}. This configuration has been extensively used in previous studies related to AMOC stability \cite{Weijer2012,DenToom2014,Brunnabend2017,VanWesten2025} and has been shown to be in reasonable agreement with observations in the subpolar North Atlantic \cite{Weijer2012,Fried2024}. The model is forced by reanalysis-derived heat and freshwater fluxes and wind stress \cite{Large2004} described in more detail in Text S1 of the Supplementary Information. This version of POP does not include interactive sea ice, and instead salinity is restored with a timescale of 30 days under a prescribed seasonal sea ice extent. No salinity restoring is applied elsewhere.

For comparison, we also use the low-resolution (1°) version of POP (LR-POP hereafter), in which mesoscale eddies are parametrized instead of explicitly captured. LR-POP has been configured to closely match the HR-POP control state (see Supplementary Information of \citeA{Weijer2012}. It uses a dipole grid with the North Pole over Greenland with 40 vertical levels. The resolution of LR-POP reaches around \SI{50}{km} in the Labrador Sea, compared to around \SI{6}{km} in HR-POP.

\subsection{Experiment setup}
We probe the impact of Greenland meltwater in the 21\textsuperscript{st} century under the SSP5-8.5 scenario \cite{ONeill2016}, the strongest global warming scenario considered in the Coupled Model Intercomparison Project phase 6 \cite<CMIP6;>{Eyring2016}. As described in Text S1 of the Supplementary Information, we derive the surface boundary conditions for ocean-only future projections from CMIP6 multi-model mean anomalies of all forcing fields except runoff. These anomalies are referenced to the 1958--2000 climatology because the POP control run was integrated using fixed ``normal-year'' forcing \cite{Large2004} representative of the 1958--2000 period. To avoid spurious breaks between historical and future forcing, we use the CMIP6 multi-model mean anomalies for both. The seasonal cycle is taken into account for all forcing anomalies, but year-to-year variations are smoothed with a 20-year running mean to isolate long-term changes and for consistency with the control run, which also has no forced interannual variability.

The historical + SSP5-8.5 (``Reference'') simulation is branched off from year 170 of an existing HR-POP control run under fixed normal-year forcing \cite{LeBars2016}. The branching point is defined as year 1978, where the global mean CMIP6-derived anomalies are closest to zero. A second simulation (``Meltwater'') is branched off from the ``Reference'' simulation in 2000 with the same forcing anomalies, but with added Greenland meltwater from the fully coupled climate--ice sheet model CESM2-CISM2 \cite{Muntjewerf2020}. The meltwater forcing is described in more detail in Text S2 of the Supplementary Information and in \citeA{Mehling2026}, who applied the same meltwater forcing in a coupled climate model. Greenland runoff is aggregated into seven drainage basins \cite{Mouginot2019} before being inserted in the uppermost model layer in a radius of around 30 km of the coast (Supplementary Fig. S1). The total Greenland runoff anomaly reaches around \SI{0.09}{Sv} in 2100.

To assess the sensitivity of our results to horizontal resolution, we perform the same set of experiments with LR-POP at 1° resolution. In addition, we test the state-dependence of Greenland meltwater impacts on the AMOC by adding the same time-dependent Greenland meltwater input as in the future projections but with the fixed ``normal-year'' forcing, mentioned earlier. Due to the computational cost of HR-POP, this experiment could only be performed with LR-POP. However, a set of more idealized experiments with HR-POP, in which a constant \SI{0.1}{Sv} runoff was inserted around Greenland with a similar geographical distribution under normal-year forcing, is available from \citeA{Weijer2012}. The cumulative meltwater input in the ``Meltwater'' simulations until 2100 is \SI{2.5}{Sv\,year}, a value reached after 25 years in the ``\SI{0.1}{Sv}'' simulations. Therefore, we compare their AMOC anomalies after 25 years (subtracting AMOC changes in the control run during the same period) to add robustness to our analysis of state-dependence.

\subsection{AMOC analysis}
The AMOC strength in the model simulations is first computed from the annual mean overturning streamfunction in depth coordinates by identifying the streamfunction maximum below \SI{500}{m} at 26.5°N. To identify drivers of AMOC changes, we relate the Atlantic overturning to the vertical structure of the meridional density gradient via the thermal wind balance relation. \citeA{Bonan2025} showed that 21\textsuperscript{st}-century AMOC weakening can be reconstructed across different CMIP6 models and scenarios using the relation
\begin{equation}
    \Psi = C \frac{g}{\rho_0 f} \langle \Delta_y \rho \rangle H^2,
    \label{eq:twb-bonan}
\end{equation}
where $\Psi$ is the AMOC strength (here defined as the maximum below \SI{500}{m} and between the equator an 30°N), $\langle \Delta_y \rho \rangle$ is the vertically integrated density difference between the North and the South Atlantic, $H$ is the scale (or pycnocline) depth defined below, $g=\SI{9.81}{\meter\per\second\squared}$ is the gravitational acceleration, $\rho_0=\SI{1026}{\kilogram\per\cubic\meter}$ is the seawater density, $f=10^{-4}$ is the Coriolis parameter at mid-latitudes, and $C$ is a proportionality constant that relates the zonal to the meridional density gradient \cite<e.g.,>{Marotzke1997,Vanderborght2025}. \citeA{Bonan2025} used $C=0.5$ but the value of $C$ is poorly constrained by theory \cite<e.g.,>{Gnanadesikan1999,Mehling2026a}. Instead, as in \citeA{Vanderborght2025} and \citeA{Mehling2026a}, we treat $C$ as a free parameter that is obtained by regressing the anomalies of the right-hand side of Eq. \eqref{eq:twb-bonan} onto the modeled AMOC anomaly for the ``Reference'' experiment, separately for the high-resolution ($C\approx1$) and low-resolution ($C\approx0.25$) versions.

Following \citeA{Bonan2025}, the density gradient $\Delta_y\rho$ is defined as the difference between the vertically averaged potential density over the upper \SI{2000}{m} between a box in the North Atlantic (43°N--65°N) and the South Atlantic (30°S--30°N). Here, the best reconstruction is achieved when the potential density is referenced to \SI{2000}{m} for HR-POP and to the surface for LR-POP. Our conclusions remain unchanged when referencing potential density to \SI{2000}{m} in both configurations (as in \citeA{Bonan2025}), but the residual would increase for LR-POP especially under ``normal-year'' forcing. The scale depth $H$ is defined, following Eq. 8 of \citeA{DeBoer2010}, as:
\begin{equation}
    \int_H^0 \Delta_y\rho(z) \:\mathrm{d}z = \frac{1}{D} \int_D^0 \int_z^0 \Delta_y\rho(z^\prime) \:\mathrm{d}z^\prime \:\mathrm{d}z.
    \label{eq:scale-depth}
\end{equation}

\section{Results}

\subsection{AMOC response to Greenland meltwater across resolutions and climate states}

\begin{figure}[htbp]
    \centering
    \includegraphics[width=0.85\linewidth]{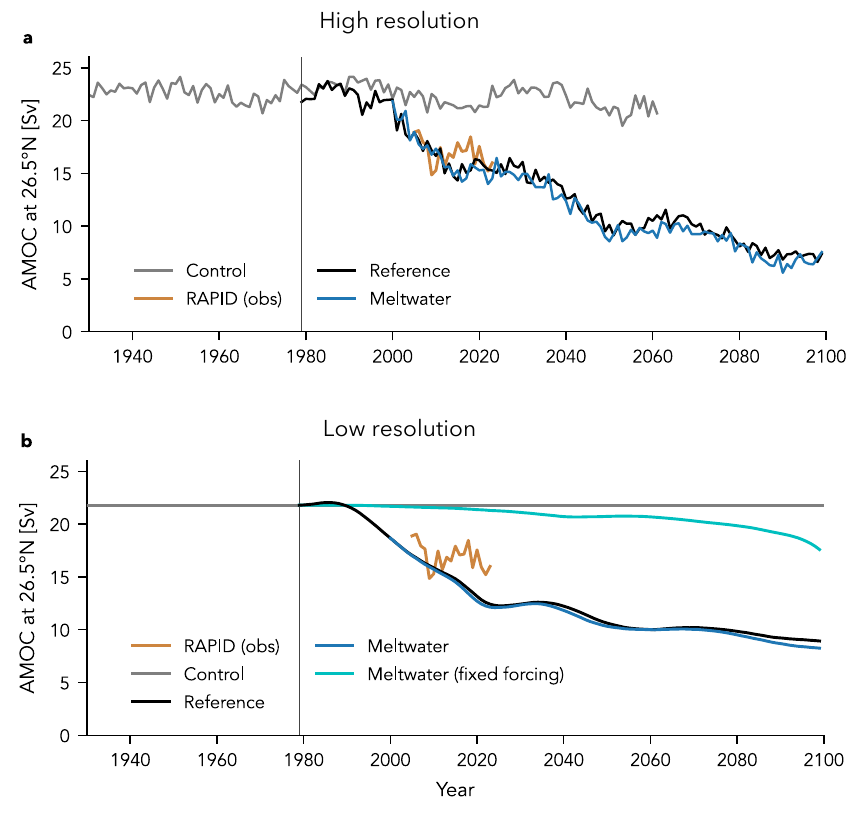}
    \caption{\textbf{AMOC timeseries at 26.5°N} under historical and SSP5-8.5 forcing in the (a) high-resolution (1/10°) POP and (b) low-resolution (1°) POP. Black lines are the ``Reference'' simulations with climate change forcing only and blue lines are the ``Meltwater'' simulations with climate change and Greenland meltwater forcing. The cyan line in panel b is with Greenland meltwater forcing only and otherwise fixed climate boundary conditions. The observed (2004--2024) AMOC time series is from the RAPID array \cite{Moat2025}. Note that interannual to multidecadal variability in the HR-POP time series arises from eddy activity \cite{Gregorio2015,LeBars2016,Leroux2018}, which is absent in LR-POP due to the low resolution.}
    \label{fig:amoc-highres}
\end{figure}

AMOC timeseries at 26.5°N are shown in Fig. \ref{fig:amoc-highres}a for the high-resolution POP experiments. The present-day (2004--2022) AMOC strength at this latitude compares favorably to the around \SI{17}{Sv} observed at the RAPID array \cite{Johns2023}. As expected, the AMOC weakens in the future projections due to the strong CO$_2$-induced climate change forcing. By the end of the century (2090--2100), the AMOC decrease compared to the present-day observational period is \SI{9.7}{Sv}, which is towards the upper end of the range of AMOC weakening simulated by CMIP6 models \cite{Weijer2020}. However, both the initial and the end-of-the-century AMOC states fall well within the CMIP6 range. The relatively monotonic AMOC weakening appears to be superimposed by intrinsic multi-decadal variability, which has previously been described in the control simulation \cite{LeBars2016}.

The effect of Greenland meltwater, i.e., the difference between the AMOC strength in the ``Meltwater'' and ``Reference'' simulations, is relatively small throughout the 21\textsuperscript{st} century. The mean difference over 2080--2100 is only \SI{0.6(2)}{Sv} but statistically significant ($p=0.003$ using a Student's $t$-test). The effect is very similar in LR-POP (\SI{0.5(1)}{Sv}, $p<0.001$), although the remaining AMOC is slightly stronger at low resolution (Fig. \ref{fig:amoc-highres}b).

While the AMOC response to Greenland meltwater at 26°N shows very little sensitivity to resolution, there is some resolution dependence when evaluating the AMOC at subpolar latitudes and in density space. In HR-POP, the largest meltwater-induced AMOC weakening at the end of the 21\textsuperscript{st} century can be found between 50°N to 70°N (Supplementary Fig. S2), while the response is relatively uniform across latitudes in LR-POP (Supplementary Fig. S3). At the location of the OSNAP observational array \cite{Fu2023} at around 60°N, the meltwater response averaged over 2080--2100 is \SI{0.9(3)}{Sv} ($p=0.001$) in HR-POP, about 50\% larger than at 26°N, but only \SI{0.5(2)}{Sv} ($p=0.014$) in LR-POP, comparable to the response at 26°N (Supplementary Fig. S4a+c). The differences can be attributed to the differing meltwater response at OSNAP-East, which accounts for most of the meltwater-induced AMOC weakening in HR-POP but changes only little in LR-POP (Supplementary Fig. S4b+d). Nevertheless, this difference does not seem to substantially impact the AMOC further south.

In contrast, large differences in the meltwater response arise when comparing different background states. When the same time-dependent Greenland meltwater input is inserted under fixed normal-year (i.e., 20\textsuperscript{th}-century) forcing instead of 21\textsuperscript{st}-century climate change, the AMOC at 26°N weakens by \SI{2.8}{Sv} in LR-POP, a more than five-fold stronger meltwater effect on the AMOC than under climate change (Fig. \ref{fig:amoc-highres}b). This response is comparable to the AMOC weakening after 25 years in the \SI{0.1}{Sv} Greenland hosing experiments of \citeA{Weijer2012} in both LR-POP (\SI{2.5}{Sv}) and HR-POP (\SI{3.0}{Sv}), indicating that the state-dependent AMOC response to Greenland meltwater input is robust across resolutions.

\subsection{Meltwater propagation pathways}
The propagation of Greenland meltwater can be compared across resolutions and background climate states using a passive freshwater tracer \cite<cf.>[]{Boning2016,Martin2022}, which follows the same model physics as the prognostic salinity. Maps of tracer concentrations by depth at the end of the 21\textsuperscript{st} century are shown in the Supplementary Material (Supplementary Fig. S5). Except for the region of meltwater input, the highest concentrations of the meltwater tracer can be found near the surface in Baffin Bay and along the Labrador Current at both resolutions. In contrast, again at both high and low resolution, much less meltwater reaches the interior of the subpolar gyre, with higher concentrations in the subtropical gyre. This similarity is despite faster and narrower boundary currents in the subpolar North Atlantic in HR-POP (Supplementary Fig. S6, cf. \citeA{Marzocchi2015,Martin2023}). The main differences between resolutions are that more meltwater remains around the Grand Banks or along North American East coast at high resolution, and that more meltwater reaches subtropical gyre, especially below 200m. Both these differences were also found by \citeA{Martin2023} with a different ocean model under historical forcing but also had little effect on the AMOC weakening.

\begin{figure}
    \centering
    \includegraphics[width=\linewidth]{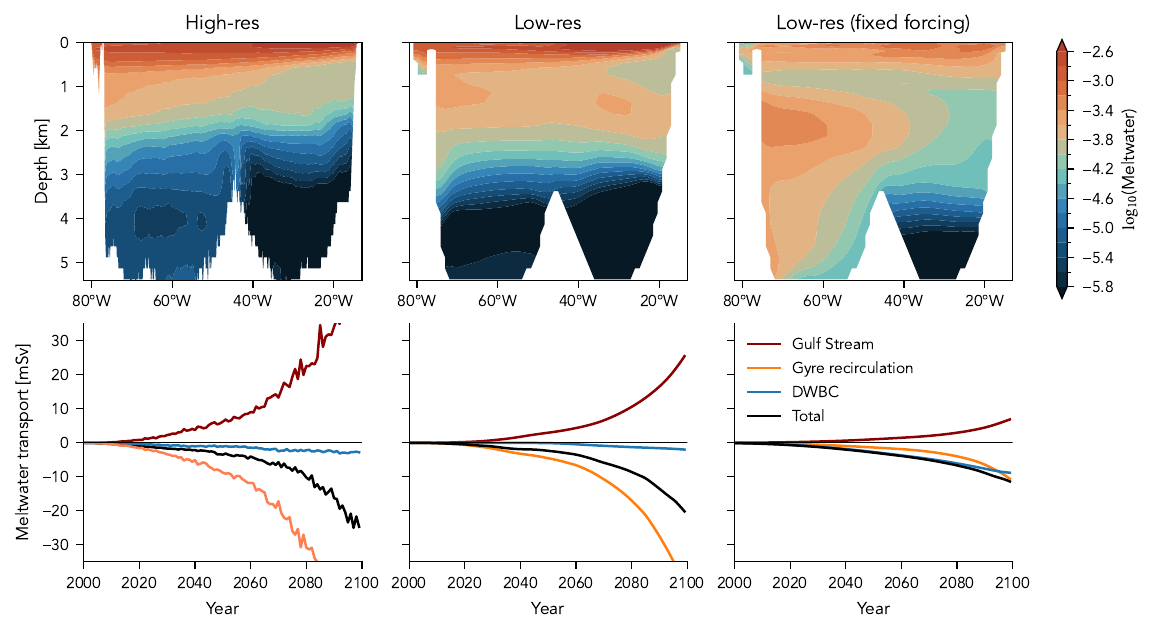}
    \caption{\textbf{Meltwater tracer concentration and transport at 26°N}: (a-c) Concentration of the meltwater tracer (on a logarithmic scale) in the year 2100, (d-f) Meltwater transport across 26°N (black) decomposed into transport by the Gulf Stream, transport in the upper \SI{1000}{m} excluding the Gulf Stream (``gyre recirculation''), and by the deep western boundary current (DWBC). The transport below \SI{1000}{m} outside of the DWBC is very small and not shown. }
    \label{fig:section-26n}
\end{figure}

To relate the meltwater propagation pathways to the AMOC, we analyze a cross-section at 26°N (Fig. \ref{fig:section-26n}) where the meridional velocity that underlies the AMOC calculation can be decomposed into components from the Gulf Stream, the deep western boundary current (DWBC), and recirculation within by the subtropical gyre \cite{Asbjornsen2023}. Following the methodology detailed in \citeA{Asbjornsen2023}, the cores of the (northward, upper-ocean) Florida and Antilles Currents, which add up to the Gulf Stream transport, as well as of the (southward, deep-ocean) DWBC are identified in the meridional velocity cross-sections. Tracer transports are then calculated by masking these cores, and the residual transport over the upper \SI{1000}{m} is defined as the subtropical gyre recirculation. At the end of the 21\textsuperscript{st} century, most of the net (southward) meltwater transport can be attributed to the upper \SI{1000}{m}, with both a large recirculation within the subpolar gyre and a southward transport mostly along the Eastern boundary. In contrast, the DWBC only accounts for around 10\% of the net meltwater transport. This fundamentally differs from the simulation under fixed normal-year forcing, where the net meltwater transport is almost identical to the meltwater transport by the DWBC, and the gyre recirculation is also much smaller.

These differences between the two climate states can be related to the different tracer distributions (first row of Fig. \ref{fig:section-26n}). In the future scenario, most of the meltwater remains near the surface at both resolutions, with highest concentrations on the Eastern side of the basin. Hence, a large portion of the southward meltwater transport appears to be via the Canary Current. Under fixed forcing, tracer concentrations in the upper ocean and especially in the Eastern North Atlantic are much lower, and in turn more meltwater has reached the deep ocean below \SI{1000}{m} (cf. Supplementary Fig. S5). This is especially the case in the DWBC on the western side of the basin, where meltwater tracer concentrations are more than an order of magnitude larger than in the future scenarios (Fig. \ref{fig:section-26n}).

\subsection{Large-scale controls of Greenland meltwater impacts}
\begin{figure}[htbp]
    \centering
    \includegraphics[width=\linewidth]{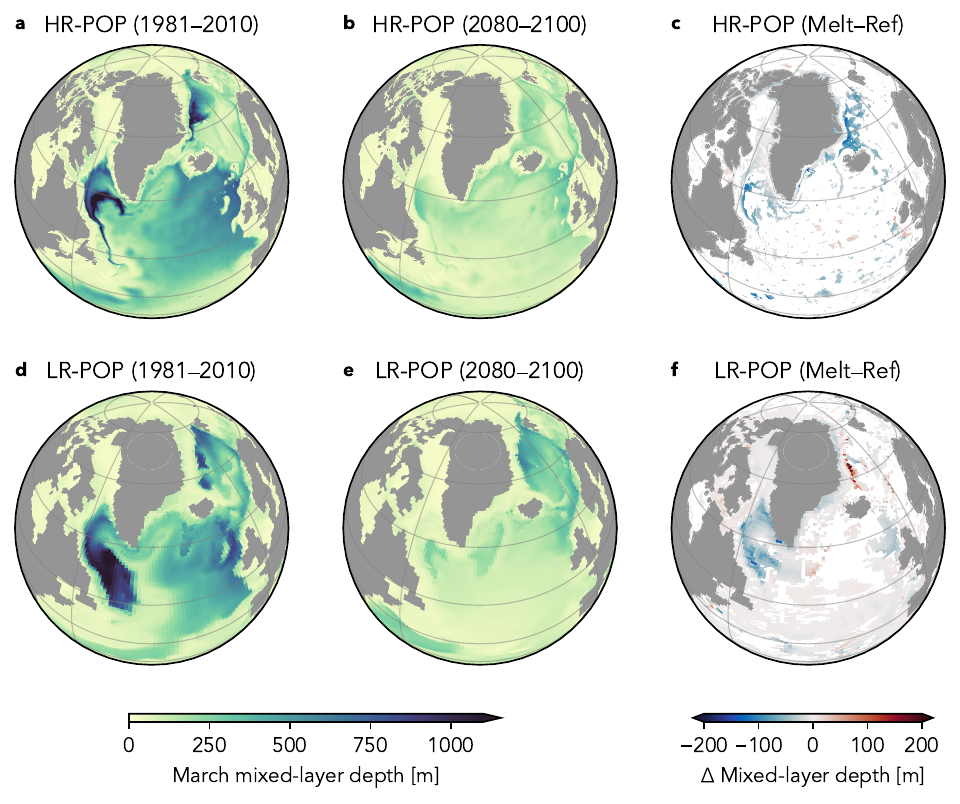}
    \caption{Mixed-layer depth in March for HR-POP (first row) and LR-POP (second row) for 1981--2010 (a, d) and 2080--2100 (b, e). For 2080--2100, mixed-layer depths are shown from the ``Meltwater'' simulations. The third column (c, f) shows the difference between the ``Meltwater'' and ``Reference'' simulations for the same period. In these panels, only grid cells with statistically significant changes ($p<0.05$ according to Welsh's $t$-test) are shown.}
    \label{fig:mld}
\end{figure}

The similar AMOC response to future Greenland meltwater input in the high- and low-resolution POP versions suggests that large-scale properties rather than (eddy-related) mesoscale features control the impacts of meltwater. One straightforward mechanism is the shallowing of deep convection under global warming, which is similar in both resolutions (Fig. \ref{fig:mld}) and linked to increasing ocean stratification \cite<cf.>{Cheng2025}. Winter mixed layers, which exceed \SI{1000}{m} in parts of the Labrador Sea and Nordic Seas in the control simulation, become shallower than \SI{500}{m} across the subpolar North Atlantic by the end of the century (Fig. \ref{fig:mld}), which provides a mechanism by which more meltwater remains confined to the upper hundreds of meters and why only very little reaches the DWBC (Fig. \ref{fig:section-26n}), independently of resolution.

Additionally, the meltwater input leads to shallower mixed layers in the subpolar gyre at both resolutions (Fig. \ref{fig:mld}c, f). Mixed layers also significantly shallow in the Nordic Seas in HR-POP but not in LR-POP, potentially explaining the stronger meltwater response at OSNAP-East in HR-POP (Supplementary Fig. S4). However, the reasons for these resolution differences north of the Greenland--Scotland ridge remain for further study, as we focus on the response at 26°N.

The AMOC response to Greenland meltwater at 26°N can be more quantitatively related to large-scale properties of the North Atlantic using the thermal wind balance framework of \citeA{Bonan2025}. Following their methodology to diagnose AMOC weakening across different future warming scenarios, we decompose AMOC changes due to global warming and/or Greenland meltwater input into a component due to changes in scale depth $H$ and meridional density gradient $\Delta \rho_y$. We compare the present-day (1981--2010) AMOC  to the AMOC state at the end of the 21\textsuperscript{st} century (2080--2100). 

\begin{figure}[htbp]
    \centering
    \includegraphics[width=\linewidth]{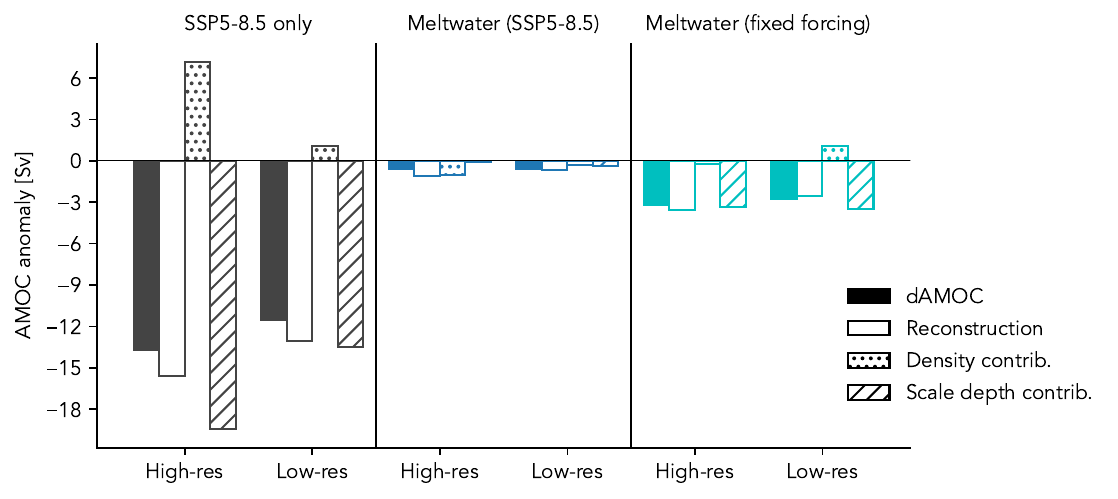}
    \caption{\textbf{AMOC changes attributed to density gradient and scale depth changes.} For each simulation, the AMOC anomaly (filled bars) is shown along with the reconstructed AMOC anomaly from thermal wind balance (non-filled bars) and its partitioning in a density gradient ($\Delta_y \rho$) and a scale depth ($H$) contribution. The pairs (high and low resolution) of AMOC anomalies are color-coded as follows. AMOC weakening (2080--2100 minus 1981--2010) due to SSP5-8.5 CO$_2$ forcing only: black, AMOC anomaly due to Greenland meltwater under SSP5-8.5 forcing (2080-2100): blue, AMOC anomaly due to Greenland meltwater under fixed normal-year forcing: cyan.}
    \label{fig:twb-contributions}
\end{figure}

As in the CMIP6 ensemble \cite{Bonan2025}, the AMOC weakening due to global warming in POP (first group in Fig. \ref{fig:twb-contributions}) is mainly driven by changes in scale depth $H$. The small contribution of density gradient changes in fact slightly opposes the AMOC weakening \cite<cf.>{DeBoer2010}. Similarly, the AMOC weakening due to Greenland meltwater under fixed climate forcing  (last group in Fig. \ref{fig:twb-contributions}) is mainly explained by the decreasing scale depth contribution. In contrast, under global warming, the Greenland meltwater input barely affects the (already strongly reduced) scale depth (second group in Fig. \ref{fig:twb-contributions}), and the meltwater effect is instead caused by a decrease in the meridional density gradient. This mechanism appears to be much less effective in weakening the AMOC than the scale depth changes. All these results are consistent across both resolutions.

One can also re-calculate the scale depth (Eq. \ref{eq:scale-depth}) for HR-POP by adding the meltwater-induced density anomalies from the ``\SI{0.1}{Sv}'' fixed-forcing simulation to the density profile for 2080--2100 under SSP5-8.5 forcing without Greenland meltwater. This leads to only a slight shallowing of the scale depth from \SI{810}{m} to \SI{804}{m}, corresponding to an AMOC weakening of \SI{0.3}{Sv}. For comparison, with a fixed background climate, the scale depth decrease induced by the same density anomaly is from \SI{1164}{m} to \SI{1126}{m}, corresponding to an AMOC weakening of \SI{3.3}{Sv}. This calculation illustrates the dominant role of mean-state changes in the ocean density profile for the impacts of Greenland meltwater under global warming.

\section{Discussion and Conclusions}
In this study, we presented the first estimate of the impacts of 21\textsuperscript{st}-century Greenland ice sheet melt on the AMOC at eddy-rich (1/10°) ocean model resolution. Our results show that the meltwater impact on AMOC strength by 2100 is statistically significant but small, between 0.5 and \SI{1}{Sv} depending on the latitude at which the AMOC is evaluated. These numbers are consistent between eddy-rich and eddy-parameterizing (1°) ocean resolutions, although the meltwater effect at the OSNAP line is slightly stronger in the eddy-rich model, potentially owing to differences in the Nordic Seas response. Our results are in good quantitative agreement with previous estimates from coupled models at 1° or coarser ocean resolution, which consistently found meltwater effects on the order of \SI{1}{Sv} until 2100 \cite{Lenaerts2015,Bakker2016,Mehling2026}.

The similar AMOC response at eddy-rich and eddy-parameterizing resolution may seem at odds with several studies claiming a potentially larger AMOC response to freshwater changes with explicit mesoscale eddies \cite{Weijer2012,Swingedouw2022,Shan2024}. However, \citeA{Weijer2012} showed that the difference between the AMOC weakening at high and low resolution is moderate as long as the meltwater input is along the Greenland coast instead of uniformly ``hosing'' the North Atlantic. In most other studies, it is difficult if not impossible to discern the role of resolution from other factors such as differences in atmosphere--ocean coupling \cite{Swingedouw2022} and different model biases at high and low resolution \cite{Shan2024}. The most comprehensive attempt to disentangle some of these factors has been by \citeA{Martin2023}, who compared the AMOC response to Greenland meltwater in high- and low-resolution configurations with and without atmosphere--ocean coupling. Our results (under global warming) confirm their finding (under fixed historical conditions) that, despite different meltwater propagation pathways, the AMOC response to Greenland runoff is similar across high- and low-resolution configurations.

To explain the physical mechanism behind the absence of a resolution-dependent AMOC response to Greenland meltwater at 26°N, we applied the AMOC reconstruction method of \citeA{Bonan2025} based on a thermal wind balance framework. This way, it was possible to relate the AMOC changes to a few large-scale metrics, in particular the scale depth $H$. This relation also appears to hold the key to the second main finding of our study, the robust state-dependence of the AMOC response to Greenland meltwater. We showed that the scale depth is barely affected when meltwater is added in a significantly warmer climate, while it controls the much stronger AMOC response to meltwater under constant historical climate conditions.

This aligns well with a previous multi-model study \cite{Swingedouw2015} which explained the state-dependence (in eddy-parameterizing coupled models) by stratification changes and differences in the ``Canary Current freshwater leakage'' \cite<cf.>{Swingedouw2013}. In our simulations, there is also a significantly enhanced southward meltwater transport along the Canary Current in a warmer climate (Fig. \ref{fig:section-26n}). \citeA{Swingedouw2015} linked this to a more zonally tilted boundary between the subtropical and subpolar gyres under global warming, which is also present in our simulations (Supplementary Fig. S7), but the exact mechanisms by which it leads to AMOC weakening remain poorly understood and should be studied in more detail in the future.

In summary, we showed that Greenland meltwater only has a moderate impact (up to around \SI{1}{Sv}) on the AMOC in future projections until 2100, even when explicitly representing mesoscale eddies, due to the significant state-dependence of meltwater input linked to stratification changes under global warming. The main limitation of our study is the use of a single model without atmosphere--ocean and ocean--sea ice coupling. Regarding sea-ice feedbacks, we implicitly account for the combined impact of global warming and AMOC weakening by prescribing a time-varying sea ice extent from CMIP6 projections, but do not capture additional feedbacks from meltwater-induced AMOC changes. However, these feedbacks are likely of second order, as previous studies showed that AMOC--sea ice feedbacks are most relevant for the transition from an ``AMOC off'' back to an ``AMOC on'' state \cite{VanWesten2024a} or very close to the AMOC tipping point \cite{Vanderborght2025}.

Regarding model-dependence, given that \citeA{Martin2023} reached similar conclusions regarding the resolution-dependence with a different ocean model, with or without atmosphere--ocean coupling, we expect that our findings would generally hold in other (coupled) models. In fact, including atmospheric coupling would be expected to further dampen the meltwater-induced AMOC weakening \cite{Martin2023}, even though mesoscale thermal feedbacks \cite{Small2008} could affect the boundary current response when coupled to a high-resolution atmospheric component. An additional caveat is that the atmospheric forcing anomalies in the SSP5-8.5 scenario were prescribed from CMIP6 models which mostly have an atmospheric resolution of around \SI{100}{km}. It has been shown that the response to warming can substantially differ at higher atmospheric resolution \cite<e.g.,>{Moreno-Chamarro2021,Wills2024}. We hope that these caveats motivate the ocean modeling community to better assess the impacts of atmospheric (forcing) resolution in the future.

The moderate impact of 21\textsuperscript{st}-century Greenland melt does not preclude stronger effects on the AMOC after 2100 \cite{Bakker2016,Mehling2026} when Greenland melt would increase further under continued warming. Nevertheless, the moderate meltwater effect contrasts with the large uncertainty in the magnitude of 21\textsuperscript{st}-century AMOC weakening due to CO$_2$ forcing alone \cite{Weijer2020,Bonan2025}. There are so far no indications that this uncertainty is linked to resolution \cite{Winton2014,Juling2021}, but instead it is often attributed to different model biases in the mean-state \cite<cf.>{Jackson2023}. Combined with these previous studies, our results therefore imply that improving (ocean) resolution alone is not a silver bullet to improving AMOC projections and that reducing coupled model biases, which can be prohibitively expensive at high resolution, might be a more promising avenue to tackling the large uncertainty in AMOC projections.

%
%

\section*{Open Research Section}
\noindent The underlying data and Jupyter Notebooks to reproduce all figures are available on Zenodo: \url{https://doi.org/10.5281/zenodo.20355645} \cite{mehling_zenodo_2026b}. RAPID data were obtained from \url{https://rapid.ac.uk/} \cite{Moat2025}.

\acknowledgments
We thank Michael Kliphuis for his kind assistance in setting up the POP simulations and for recovering data from \citeA{Weijer2012}, Elian Vanderborght for insightful discussions about the thermal wind balance framework, and Andy Hogg for valuable clarifications about the setup of \citeA{Li2023a}. We thank the climate modeling groups participating in CMIP6 for making their output freely available. We also wish to thank the two reviewers whose
comments have significantly improved the paper.

\noindent This work was funded by the European Research Council through the ERC-AdG project TAOC (PI: Dijkstra, project 101055096). The model simulations were performed on the Dutch National Supercomputer Snellius within NWO‐SURF project 2024.013.

%

\renewcommand{\refname}{Additional references from the Supporting Information}

\end{document}